\documentclass[
prd,aps,twocolumn, showpacs,tightenlines,nofootinbib,preprintnumbers, superscriptaddress
]{revtex4}
\usepackage{graphicx}
\usepackage{amssymb,amsmath,latexsym}
\usepackage{amsfonts}

\usepackage{cancel}
\usepackage{color}
\input{colordvi.tex}
\input{epsf}
\usepackage{epsf}
\usepackage{graphicx,epsfig}
\usepackage{bm}
\usepackage{latexsym,float}

\newcommand{\bear}{\begin{array}}  \newcommand{\eear}{\end{array}}
\newcommand{\beq}{\begin{equation}}  \newcommand{\eeq}{\end{equation}}
\newcommand{\bef}{\begin{figure}}  \newcommand{\eef}{\end{figure}}
\newcommand{\bec}{\begin{center}}  \newcommand{\eec}{\end{center}}

\newcommand{\beqa}{\begin{eqnarray}}
\newcommand{\eeqa}{\end{eqnarray}}
\newcommand{\p}{\phi}

\newcommand{\siml}{\lesssim}

\newcommand{\vp}{\varphi}
\newcommand{\Om}{\Omega}
\newcommand{\Omp}{\Omega_{\phi 0}}

\newcommand{\pii}{\phi_i}

\newcommand {\ga} {\ {\raise-.5ex\hbox{$\buildrel>\over\sim$}}\ }
\newcommand {\la} {\ {\raise-.5ex\hbox{$\buildrel<\over\sim$}}\ } 
\newcommand{\eqn}[1] {Eq.~(\ref{#1})}
\newcommand{\fig}[1] {Fig.~(\ref{#1})}
\newcommand{\phid}{\dot{\phi}}
\def\be{\begin{equation}}
\def\ee{\end{equation}}
\def\ba{\begin{eqnarray}}
\def\ea{\end{eqnarray}}
\renewcommand{\(}{\left(} 
\renewcommand{\)}{\right)} 
\renewcommand{\[}{\left[} 
\renewcommand{\]}{\right]} 

\begin{document}

\title{Slow-roll k-essence}

\author{Takeshi Chiba}
\email{chiba@phys.chs.nihon-u.ac.jp}
\affiliation{Department of Physics, \\
College of Humanities and Sciences, \\
Nihon University, \\
Tokyo 156-8550, Japan}

\author{Sourish Dutta}
\email{sourish.d@gmail.com}
\affiliation{Department of Physics and Astronomy,\\
Vanderbilt University, \\
Nashville, TN 37235}

\author{Robert~J. Scherrer}
\email{robert.scherrer@vanderbilt.edu}
\affiliation{Department of Physics and Astronomy,\\
Vanderbilt University, \\
Nashville, TN 37235}

\date{\today}


\begin{abstract}
We derive slow-roll conditions for thawing k-essence with a separable Lagrangian $p\(X,\phi\)=F\(X\)V\(\phi\)$. 
We examine the evolution of the equation of state parameter, $w$, as a function of the
scale factor $a$, for the case where $w$ is close
to $-1$.  We find two distinct cases, corresponding to $X \approx 0$ and $F_X \approx 0$, respectively. 
For the case where $X\approx0$ the evolution of $\phi$ and hence $w$ is described by only 
two parameters, and $w(a)$ is model-independent and coincides with similar behavior seen in thawing quintessence models.
This result also extends to non-separable Lagrangians where $X\approx0$.
For the case
$F_X \approx 0$, an expression is derived for $w(a)$, but this expression
depends on the potential $V(\phi)$, so there is no model-independent limiting
behavior.  For the $X \approx 0$ case, we derive observational constraints
on the two parameters of the model, $w_0$ (the present-day value of $w$), and
the $K$, which parametrizes the curvature of the potential.  We find that the
observations sharply constrain $w_0$ to be close to $-1$, but provide very poor
constraints on $K$.
\end{abstract}

\maketitle

\section{Introduction}

Cosmological data from a wide range of sources including type Ia supernovae \cite{union08, perivol, hicken}, the cosmic microwave background \cite{Komatsu}, baryon acoustic oscillations \cite{bao,percival}, cluster gas fractions \cite{Samushia2007,Ettori} and gamma ray bursts \cite{Wang,Samushia2009} seem to indicate that at 
least 70\% of the energy density in the
universe is in the form of an exotic, negative-pressure component,
called dark energy.  (See Ref. \cite{Copeland} for a recent
review).

The dark energy component is usefully parameterized by its equation of state (EoS) parameter 
$w$, defined as the ratio of its pressure to its density. 
Observations constrain $w$ to be very close to $-1$.
For example, if $w$ is assumed to be constant, then $-1.1 \la w \la -0.9$  \cite{Wood-Vasey,Davis}.

On the other hand, a variety of models have been proposed in which $w$ is time varying.
A common approach is to use a scalar field as the dark energy component.
The class of models in which the scalar field is canonical is dubbed quintessence
\cite{RatraPeebles,CaldwellDaveSteinhardt,LiddleScherrer,SteinhardtWangZlatev}
and has been extensively studied. In most of these models the field rolls slowly on a flat potential. 
It is also possible to have models in which the field is trapped in a false vacuum 
(see \cite{DuttaHsuScherrerReeb} and references therein for examples of such models).  

A related, yet somewhat different approach is phantom dark energy, i.e., a component for which 
$w<-1$, as first proposed by Caldwell \cite{Caldwell}. Such models have well-known problems
\cite{CarrollHoffmanTrodden,ClineJeonMoore,BuniyHsu,BuniyHsuMurray} (however see 
\cite{Creminelli:2008wc} for recent attempt to construct a stable model),
but nevertheless have been widely studied as potential dark energy candidates.

In both the above approaches, the potential energy of the scalar field
is responsible for bringing about the late-time acceleration of the Universe.
A third type of model utilizes the kinetic energy of the field through the use
a non-canonical kinetic term in the Lagrangian. Such fields were first used
in the context of inflation, in a scenario that is called k-inflation \cite{adm}. 
They have since been
studied as dark energy candidates \cite{coy,ams,Chiba:2002mw} and 
these models are called
``k-essence''. 

Given the considerable freedom that exists in choosing the potential
function of the scalar field $V\(\phi\)$, as well as the kinetic function $F\(X\)$ for k-essence, 
it would be interesting to find any model-independent evolution for $w(a)$. Some recent 
work has uncovered such model-independent evolution for quintessence and phantom models when 
the field evolves in the vicinity of extrema of the potential, and $w\sim-1$. 
Ref. \cite{ScherrerSen1} considered the evolution of a 
a scalar field, initially at rest, in a potential satisfying the
``slow-roll" conditions:
\ba
\label{SR1}\[\frac{1}{V}\frac{dV}{d\phi}\]^2\ll 1,\\
\label{SR2}\left|\frac{1}{V}\frac{d^2 V}{d\phi^2}\right|\ll 1.
\ea
The first condition ensures that $w$ is close to $-1$, while
the two conditions taken together indicate that $(1/V)(dV/d\phi)$ is nearly
constant.
In the terminology of Ref. \cite{CL}, these are ``thawing" models.

For all potentials satisfying these conditions,
it was shown in \cite{ScherrerSen1} that the
behavior of $w$ can be accurately described by
a unique expression depending only on the present-day
values of $\Omega_\phi$ and the initial value of $w$.
In \cite{ScherrerSen2} this result was extended to
phantom models satisfying
Eqs.~(\ref{SR1}-\ref{SR2}), and the $w$ dependence of
these phantom models was shown to be described by
the same expression as in the quintessence case. 

The slow roll conditions, Eqs.~(\ref{SR1}-\ref{SR2}),
while sufficient to ensure $w\simeq-1$ today, are not necessary.
In \cite{ds1}, a second possibility was considered,
in which equation (\ref{SR1}) holds, but equation (\ref{SR2}) is relaxed.
This corresponds to a quintessence field rolling near a local
maximum of its potential.  As in the case of slow-roll quintessence,
this case can be solved analytically.  In this case, there is
an extra degree of freedom, the value of $(1/V)(d^2 V/d\phi^2)$,
so that instead of a single solution for the evolution of $w$,
one obtains a family of solutions that
depend on the present-day
values of $\Omega_\phi$ and $w$ and the value of $(1/V)(d^2 V/d\phi^2)$
at the maximum of the potential.  This family of solutions includes
the slow-roll solution as a special case in the limit
where $(1/V)(d^2 V/d\phi^2) \rightarrow 0$. The corresponding result for phantom fields, 
where the field rolls near the minimum of its potential was derived in \cite{ds2} - and an 
identical expression for $w(a)$ was obtained. In both the above cases, it was shown that 
the expression for $w(a)$ reduces to the corresponding ones in \cite{ScherrerSen1} 
and \cite{ScherrerSen2} as the potential gets flat, i.e., $(1/V)(d^2 V/d\phi^2) \rightarrow 0$. 

Ref. \cite{ds3} examined the opposite case of
dark energy models in which a quintessence or a phantom
field rolls near the vicinity of a local minimum or
maximum, respectively, of its potential. It was shown that as long as \eqn{SR1} is satisfied, 
(although \eqn{SR2} need not be), the evolution of $w$ is described by an expression identical 
to the one in \cite{ds1,ds2}. In these cases the evolution of $w$ was found to encompass a 
richer set of behaviors, including oscillating solutions. 

In \cite{chiba}, one of us (TC) showed that the expression derived in \cite{ds1,ds2,ds3} has a wider 
applicability than the cases of fields rolling close to extrema in their potentials.  
The following more general slow-roll conditions on the potential are derived there:
\ba
\label{SR1chiba}\epsilon\equiv\frac{V'^2}{6H^2 V};&&\epsilon\ll1,\\
\label{SR2chiba}\eta\equiv\frac{V''}{3H^2};&&\vert\eta\vert\ll1,
\ea
and while retaining the assumption that $w\approx -1$, dropped the assumption that the field 
is close to a local extremum in the potential. Interestingly, once again the expression for 
the evolution of $w$ under these more general conditions was found to coincide exactly with 
the one in refs. \cite{ds1,ds2,ds3}. 

In this paper we investigate thawing models in k-essence. We derive slow-roll conditions for 
thawing k-essence analogous to equations \eqref{SR1chiba}-\eqref{SR2chiba}, and show that, 
when $w\sim-1$, in some cases one does obtain the same model independent evolution of $w$ 
seen in the above references.

\section{Slow-Roll Thawing K-Essence}

The Lagrangian density of k-essence \cite{adm,coy,ams} is $p(\p,X)$, where 
$X=-\nabla^{\mu}\p\nabla_{\mu}\p/2$. The pressure $p_{\p}$ of the scalar
field $\phi$ is given by $p(\p,X)$ itself and the energy
density $\rho_{\p}$ is given by $\rho_{\p}=2X(\partial p/\partial X)-p$
\cite{adm,coy}, so that the equation of state parameter, $w$, is
\begin{equation}
\label{genw}
w = \frac{p}{2X(\partial p/\partial X)-p}.
\end{equation}

Working in units of $8\pi G=1$, the basic equations in a flat universe are
\beqa
&&\ddot{\p}\left(\frac{\partial p}{\partial X}+
  \dot\p^{2}\frac{\partial^{2}p}{\partial X^2}\right)\nonumber\\
  &&+3H\frac{\partial
  p}{\partial X}\dot\p+\frac{\partial^{2}p}{\partial X\partial
  \p}\dot\p^{2}-\frac{\partial p}{\partial\p} = 0,
\label{eq:eom1}\\
&&H^2 = \left(\frac{\dot a}{a}\right)^2= \frac13(\rho_B+\rho_{\p})\nonumber\\
&&= \frac13\left(\rho_B+2X\frac{\partial p}{\partial
    X}-p\right), 
\label{eq:hubble} \\
&&\frac{\ddot{a}}{a} = -\frac16(\rho_B+3p_B+\rho_{\p}+3p_{\p})\nonumber\\
&&=-\frac16
\left((1+3w_B)\rho_B+(1+3w)\rho_{\p}\right),
\label{eq:a}
\eeqa
where $\rho_B$ and $p_B$ are the energy density and the pressure of
the background matter and/or radiation, respectively. 

We now proceed to derive slow-roll conditions for k-essence with 
the following factorized form of $p(\p,X)$:
\beq
p(\p,X)=V(\p)F(X).
\label{lagrangian}
\eeq
The equation of motion of the scalar field is then written as
\beq
\ddot\p\left(F_X+2XF_{XX}\right)+3HF_X\dot\p+\left(2XF_X-F\right)\frac{V'}{
  V}=0,
\label{eq:eom2}
\eeq
where $F_X=dF/dX$ and $V'=dV/d\p$. 
We also introduce the sound speed of k-essence,  which is the relevant quantity for the growth of 
density perturbations, 
\beqa
c_s^2=\frac{\partial p/\partial X}{\partial \rho/\partial X}=\frac{F_X}{2XF_{XX}+F_X}.
\label{soundspeed}
\eeqa
Using $c_s^2$, The equation of motion Eq. (\ref{eq:eom2}) is rewritten as
\beqa
\ddot\p+3c_s^2H\dot\p+c_s^2\frac{2XF_X-F}{F_X}\frac{V'}{V}=0.
\label{eomp}
\eeqa

\subsection{Slow-Roll Conditions for K-Essence}

By slow-roll k-essence, we mean a model of k-essence whose equation of state $w$ is close 
to $-1$ so that
\beqa
|XF_X|\ll |F|.
\label{slowroll1}
\eeqa
Thawing models correspond to the equation of state $w=p_{\p}/\rho_{\p}$ very close to $-1$, 
so that the Hubble friction is not effective and hence $\ddot\p$ is not necessarily small compared 
with $3H\dot\p$ in Eq. (\ref{eomp}).  

We derive the slow-roll conditions for thawing k-essence during the
matter/radiation 
dominated epoch. 
Generalizing the corresponding expression for quintessence
\cite{chiba,Linder,Crittenden}, we first introduce the following function:
\beqa
\beta=\frac{\ddot\p}{3c_s^2H\dot\p}.
\label{beta}
\eeqa
As stated above, for thawing models, $\beta$ is a quantity of  ${\cal O}(1)$. We assume 
$\beta$ is approximately constant in the sense that $|\dot\beta|\ll H|\beta|$, and 
the consistency of this assumption 
will be checked later. In terms of $\beta$, from Eq. (\ref{eomp}) using Eq. (\ref{slowroll1}), 
$\dot\p$ is written as
\beqa
\dot\p =\frac{FV'}{3(1+\beta)HF_XV},
\label{pdot}
\eeqa
and the slow-roll condition Eq. (\ref{slowroll1}) becomes
\beqa
\epsilon=\frac{|F|V'^2}{6H^2|F_X|V^2}\ll 1, 
\label{slow:cond:1}
\eeqa
where we have omitted $1+\beta$ since it is an ${\cal O}(1)$ quantity, and we
have
introduced the factor of 
$1/6$ so that $\epsilon$ coincides with the slow-roll parameter for thawing quintessence \cite{chiba}, 
$\epsilon=\frac16(V'^2/H^2V)$. \footnote{A canonically normalized scalar field $\vp$ 
corresponds to $d\vp=V^{1/2}d\p$ for $F=X-1$.} 
Eq. (\ref{slow:cond:1}) is a k-essence counterpart of the quintessence 
slow-roll condition $V'^2/H^2V\ll 1$. 

Similar to the case of inflation, the consistency of Eq. (\ref{beta}) and Eq. (\ref{eomp}) 
should give the second slow-roll condition. In fact, from the time derivative of Eq. (\ref{pdot}) 
we obtain 
\beqa
\beta=\left(\frac{V''}{V}-\frac{V'^2}{V^2}\right)\frac{F}{9(1+\beta)H^2F_X}
+\frac{(1+w_B)}{2},
\label{betaeq}
\eeqa
where we have used $\dot H/H^2\simeq -3(1+w_B)/2$ from Eqs. (\ref{eq:hubble})-(\ref{eq:a}) 
and have assumed $\dot\beta\ll H\beta$. While  
the left-hand-side of Eq. (\ref{betaeq}) is an almost time-independent quantity by assumption, 
the first term in the right-hand-side is 
a time-dependent quantity in general. Therefore the equality holds if 
the first term is negligible, which requires in addition to Eq. (\ref{slow:cond:1})
\beqa
\eta=-\frac{FV''}{3H^2F_XV}; ~~~~~~~~|\eta|\ll 1,
\label{slow:cond:2}
\eeqa
so that $\beta$ becomes
\beqa
\beta=\frac{1+w_B}{2},
\label{betasol}
\eeqa
{\it or} $\eta$ itself becomes a constant so that 
\beqa
\eta= -3(1+\beta)\left(\beta-\frac12 (1+w_B)\right).
\label{etasol}
\eeqa
The former condition would correspond to the slow-roll models with $X\simeq 0$, 
while the latter corresponds to 
the slow-roll models with $F_X\simeq 0$.  The expression for
$\beta$ given by Eq. (\ref{betasol}) is approximately constant, which is  consistent 
with our assumption. 
Here the factor $-1/3$ is introduced in Eq. (\ref{slow:cond:2}) so that $\eta$ coincides with 
the slow-roll parameter for thawing quintessence \cite{chiba}, $\eta=\frac13 (V''/H^2)$. 
\footnote{In terms of a canonically normalized scalar field $\vp$, $3\eta H^2=V_{,\vp\vp}+
\frac12 V_{,\vp}^2/V$, and when combined with $\epsilon\ll 1$, $\eta \simeq 
\frac13 (V_{,\vp\vp}/H^2)$.} 
Eq. (\ref{slow:cond:2}) is a k-essence counterpart of the quintessence slow-roll condition 
$|V''|/H^2\ll 1$.

Eq. (\ref{slow:cond:1}) and Eq. (\ref{slow:cond:2}) constitute the slow-roll conditions for thawing 
k-essence during the matter/radiation epoch.  Note that these expressions have assumed a negligible
contribution to the expansion rate from the k-essence itself, 
and so $\beta$ is no longer a constant and Eq. (\ref{betasol}) or Eq. (\ref{etasol}) 
becomes progressively less accurate 
as the k-essence begins to dominate at late times.  In what follows, we do
{\it not} make the assumption of matter/radiation domination, so that our results will be accurate up
to the present.

\subsection{Parametrizing the Equation of State}

Next we derive general solutions for $\p$ in the limit where $|1+w|\ll 1$, and we derive $w$ as a 
function of $a$. We note that $1+w=0$ implies (a) $X=0$ or (b) $F_X=0$. In the following we consider 
each case. 

\paragraph*{Case (a):}
First we consider the case where $X\simeq 0$. In this case, $c_s^2 \approx 1$, and Eq. (\ref{eomp})
simplifies to
\beqa 
\ddot\p+3H\dot\p-\frac{F(0)V'}{F_X(0)V}=0.
\eeqa
The Hubble friction term in Eq. (\ref{eomp}) can be eliminated by the 
following change of variable \cite{ds1}
\beqa
u=(\p-\p_i) a^{3/2},
\eeqa
where $\p_i$ is an arbitrary constant,  which is introduced for later use, 
and then Eq. (\ref{eomp})  becomes
\beqa
\ddot u+\frac34 (p_B+p_{\p}) u-a^{3/2} \frac{F(0)V'}{F_X(0)V}=0.
\label{eomu}
\eeqa
We assume a universe consisting of matter and k-essence with $w\simeq -1$. 
Then the pressure is well approximated by a constant: $p_B+p_{\p}\simeq p_{\p}\simeq -\rho_{\p 0}$, 
where $\rho_{\p 0}$ is the nearly constant density contributed by the k-essence 
in the limit $w\simeq -1$. Eq. (\ref{eomu}) then becomes
\beqa
\ddot u-\frac34 \rho_{\p 0} u+a^{3/2}  \frac{F(0)V'}{F_X(0)V}=0. 
\label{eomu2}
\eeqa
Since we consider a slow-roll scalar field ($X\simeq 0$),  
the potential may be generally expanded around 
some value $\p_i$, which we identify with the initial value, in the form (up to quadratic order)
\be
\begin{split}
V(\p)=V(\p_i)+V'(\p_i)(\p-\p_i)+\\
\frac12V''(\p_i)(\p-\p_i)^2.
\label{exp}
\end{split}
\ee
Substituting the expansion given by Eq. (\ref{exp}) into Eq. (\ref{eomu2}) and taking 
$\rho_{\p 0}=-F(0)V(\pii)$  gives
\be
\begin{split}
\ddot u+\left(-\frac{F(0)V''(\pii)}{F_X(0)V(\pii)}+\frac34 F(0)V(\pii)\right)u=\\
\frac{F(0)V'(\pii)}{F_X(0)V(\pii)}a^{3/2}.
\label{eomu3}
\end{split}
\ee
This equation is identical to Eq. (19) in Ref. \cite{chiba} by the substitution:
$V\rightarrow -FV, V''\rightarrow -FV''/F_XV, V'\rightarrow -FV'/F_XV$. Therefore, 
the evolution of $\phi$ is the same (in functional form) and the equation of state 
is again given by the same functional form derived in \cite{ds1,chiba}: 
\begin{widetext}
\beqa
\label{finalfinal}
1+w(a)=(1+w_0)a^{3(K-1)}\left(\frac{(K-F(a))(F(a)+1)^K+(K+F(a))(F(a)-1)^K}
{(K-\Omp^{-1/2})(\Omp^{-1/2}+1)^K+(K+\Omp^{-1/2})(\Omp^{-1/2}-1)^K}\right)^2,
\eeqa
\end{widetext}
where $K$ and $F(a)$ [not to be confused with $F(X)$] are defined by 
\beqa
K=\sqrt{1-\frac43\frac{V''(\pii)}{F_X(0)V(\pii)^2}},\\
F(a)=\sqrt{1+(\Omp^{-1}-1)a^{-3}}.
\eeqa

This equivalence with a scalar field having a canonical kinetic term can also be 
easily seen by noting that $F(X)$ can be expanded (if it is analytic at $X=0$) 
for small $X$ as $F(X)=F(0)+F_X(0)X$ and the Lagrangian reduces to that of a canonical scalar field 
by field redefinition. This implies that the equivalence is not limited to the 
factorized form of $p(\p,X)$ (Eq. \ref{lagrangian}), but holds for more general $p(\p,X)$ 
because $p(\p,X)$ can be expanded (if it is analytic at $X=0$) 
for small $X$ as $p(\p,X)=p(\p,0)+p_X(\p,0)X$.

\paragraph*{Case (b):} 
For the case of $F_X\simeq 0$, we may expand $F(X)$ around the extremum of $F$ (say at $X=X_m$) 
as (like hilltop quintessence)
\beqa
F(X)=F(X_m)+\frac12 F_{XX}(X_m)(X-X_m)^2.
\eeqa
In this case, assuming $X\simeq X_m$, Eq. (\ref{eomp}) simplifies to
\beqa 
\dot X +3H(X-X_m)-\frac{F(X_m)}{2X_mF_{XX}(X_m)}\frac{d}{dt}(\ln V)=0.
\eeqa
\\
Therefore, the evolution of $X$ depends on the shape of $V(\phi)$, which is no longer 
Taylor expanded because $\phi$ can evolve significantly in this case.

However, one can still derive an analytic expression for $w(X)$ . Starting from \eqn{genw}, one can write $w(X)$ as
\be
w(X)=\frac{F(X)}{2XF_X\(X\)-F(X)}.
\ee
For $F_X(X)\rightarrow 0$, this expression can be expanded
about the point $X=X_m$. In this case terms up to second order must be retained,
because while
$(X-X_m)^2$ is small, the second Taylor coefficient can be large
leading to a large contribution.
The resulting expansion gives:
\begin{widetext}
\ba
\label{Xexpand}
w\(X\)&=&-1-\[\frac{2 X_m
F_{XX}(X_m)}{F(X_m)}\](X-X_m)-\[\frac{2F_{XX}(X_m)}{F(X_m)}+\frac{4
X_m^2 F_{XX}^2(X_m)}{F^2(X_m)}\]\(X-X_m\)^2 .
\ea
\end{widetext}
Our numerical results indicate that equation (\ref{Xexpand}) is an excellent approximation for the
evolution of $w$ for the case where $F_X \rightarrow 0$ (see Fig. 4 below).  However, it is of 
limited usefulness, since $w(a)$ in this case depends on $X(a)$, and $X(a)$, in turn, 
depends on the functional form of $V(\phi)$.

\section{Comparison to numerical results}

We now turn to numerically solving the equations of motion in order to compare \eqn{finalfinal} 
against the exact evolution. We consider a Universe consisting of perfect fluid dark matter and 
k-essence dynamical dark energy. 

\begin{figure}
	\epsfig{file=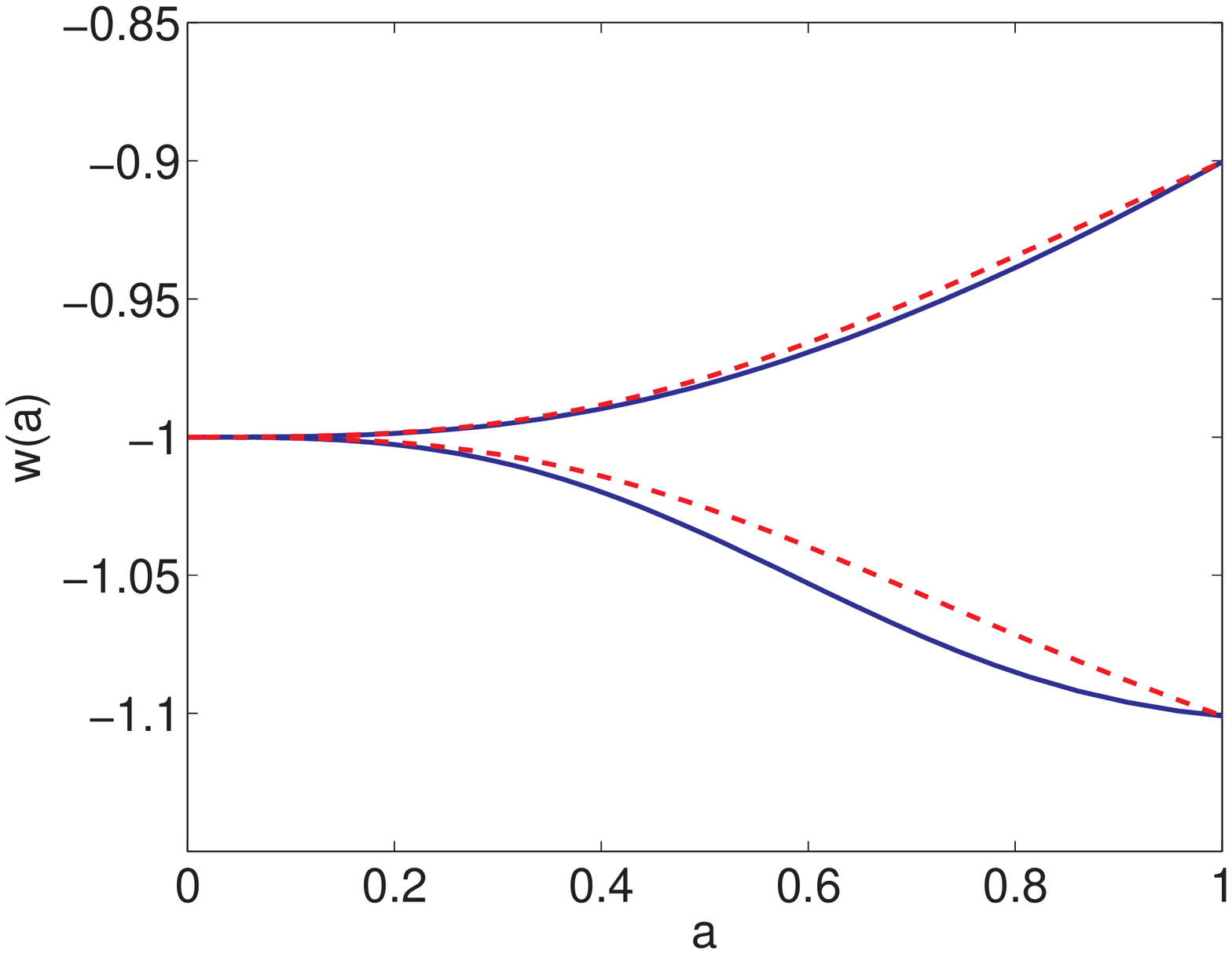,height=55mm}
	\caption
	{	\label{gauss} Evolution of $w(a)$ for a k-essence model with $F\(X\)=\sqrt{1-mX}$ and  $V\(\phi\)=A
	e^{-\phi^2/\sigma^2}$. The upper plot corresponds to the $m=1$ case and the lower one corresponds
	to the $m=-1$ case. The solid (blue) lines denote the numerical result and the broken (red) lines
	denote the analytic approximation given by \eqn{finalfinal}.}
\end{figure}

\begin{figure}
	\epsfig{file=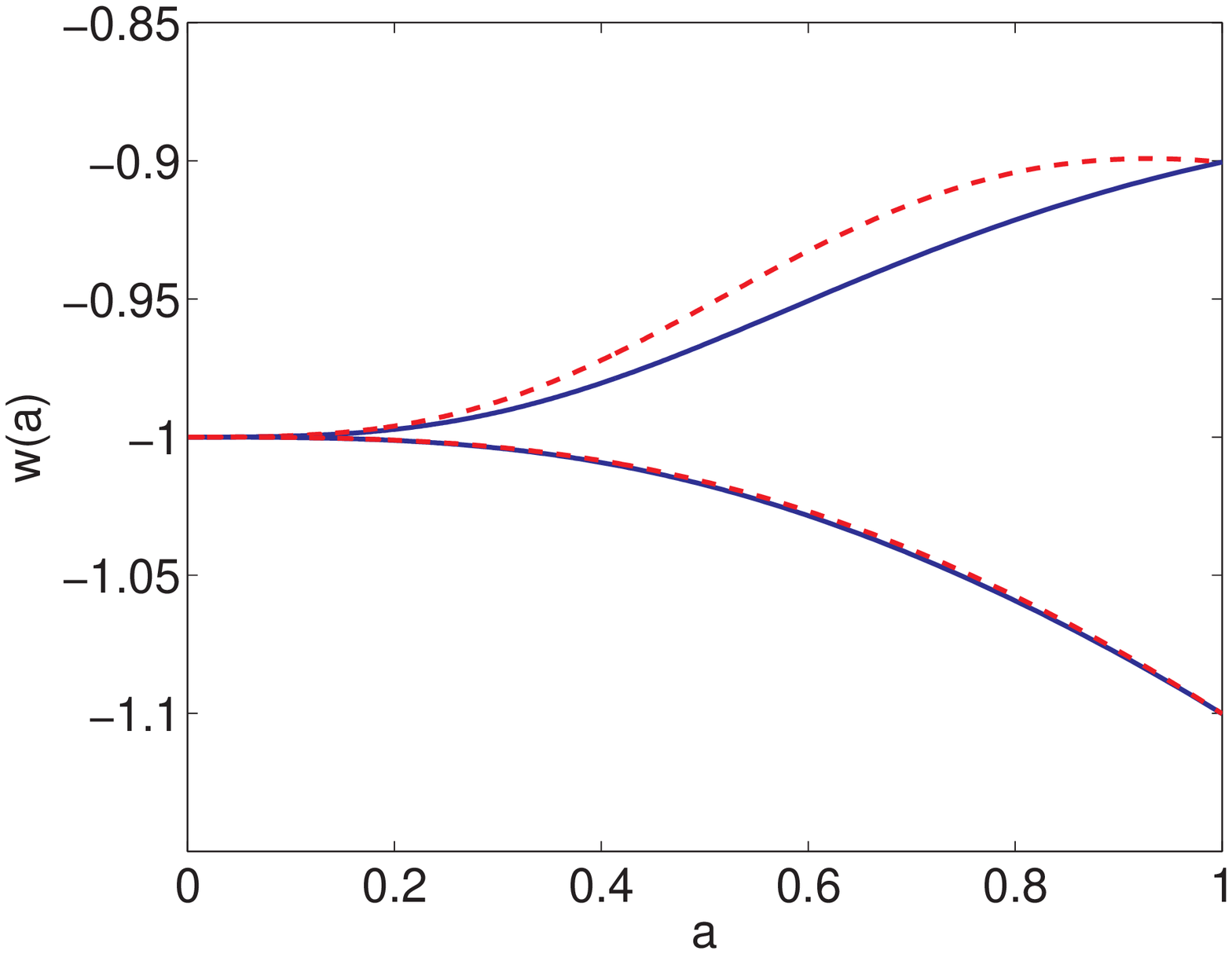,height=55mm}
	\caption
	{	\label{powerlaw} Evolution of $w(a)$ for a k-essence model with $F\(X\)=\sqrt{1-mX}$,
	$V\(\phi\)=A\phi^{-\alpha}$.  The upper plot corresponds to the $m=1$ case and the lower one
	corresponds to the $m=-1$ case. The solid (blue) lines denote the numerical result and the broken
	(red) lines denote the analytic approximation given by \eqn{finalfinal}.}
\end{figure}

\begin{figure}
	\epsfig{file=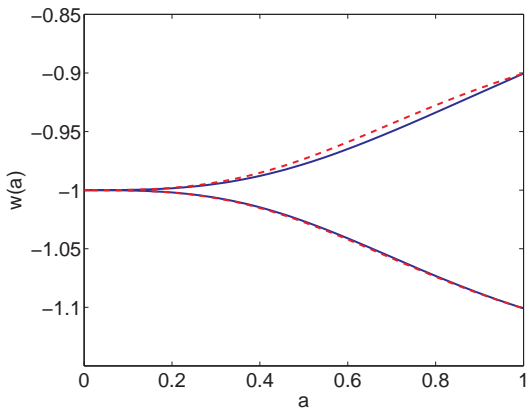,height=55mm}
	\caption
	{	\label{dbi} Evolution of $w(a)$ for a DBI model with
	$p\(\phi,X\)=mT(\phi)-mT(\phi)(1-2X/T(\phi))^{1/2}-V(\phi)$, $T\(\phi\)=\phi^4$,
	$V\(\phi\)=\mu^2\phi^2$.  The upper plot corresponds to the $m=1$ case and the lower one corresponds
	to the $m=-1$ case. The solid (blue) lines denote the numerical result and the broken (red) lines
	denote the analytic approximation given by \eqn{finalfinal}.}
\end{figure}

We work with models which satisfy the slow roll conditions derived above, i.e., \eqn{slow:cond:1} 
and \eqn{slow:cond:2}. For the cases where $X\approx 0$, the three specific models considered are 
listed below. 
\begin{enumerate}
	\item{Case 1}: \\$F\(X\)=\sqrt{1-mX}$, $V\(\phi\)=A e^{-\phi^2/\sigma^2}$, see \fig{gauss}. 
      This is the rolling tachyon Lagrangian \cite{sen} suggested by the boundary 
      string field theory \cite{kmm}.
	\item{Case 2}: \\$F\(X\)=\sqrt{1-mX}$, $V\(\phi\)=A\phi^{-\alpha}$, see \fig{powerlaw}. 
      This is the model studied in \cite{ass}. 
	\item{Case 3}: \\$p\(\phi,X\)=mT(\phi)-mT(\phi)(1-2X/T(\phi))^{1/2}-V(\phi)$, $T\(\phi\)=\phi^4$,
	$V\(\phi\)=\mu^2\phi^2$ see \fig{dbi}.  This is the Dirac-Born-Infeld (DBI) model discussed 
in Refs. \cite{Martin:2008xw,Ahn:2009hu}
\end{enumerate}
where $m=\pm1$ for all three cases.
In the first two cases, we choose $\phi(t=t_i)=1$ and $\phid(t=t_i)=0$.
The constants in the potentials ($A$, $\sigma$ and $\alpha$) are then adjusted to give 
$\rho_\phi=\rho_{\Lambda}$ and $w=-1$ at $t=t_i$ and $w=-0.9$ or $w=-1.1$ at $t=t_0$. 
For the third case, we choose $\phid(t=t_i)=0$ and adjust $\phi_i$ and $\mu$ to get 
the above initial and final conditions.

In all these cases we find excellent agreement between the numerical and analytic 
results, i.e., $\delta w/w\leq 0.01$. The success of our approximation for the DBI case 
indicates that, as noted earlier, it is not just restricted to separable Lagrangians. 
Note that for the second case it is found in Ref.\cite{ass} that  models 
have a unique equation of state, which corresponds to $K=1$ 
in Eq. (\ref{finalfinal}). 

\begin{figure}
	\epsfig{file=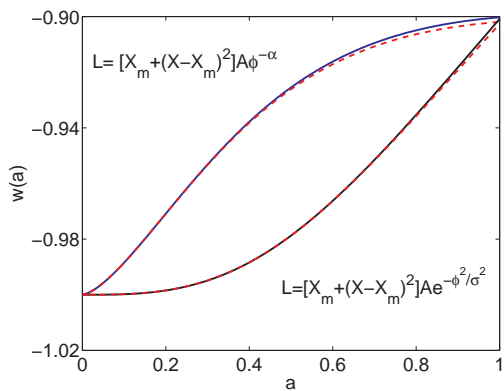,height=55mm}
	\caption
	{	\label{F_X} Solid lines give evolution of $w(a)$ for the models
	indicated, for both of which $F_X\approx 0$ throughout the evolution.  Broken
	lines give the analytic approximation.  Both the exact evolution and the analytic approximation
	depend on the choice of $V(\phi)$, as shown.}
\end{figure}

As noted earlier, for the set of models where $F_X\approx 0$, we do not expect
the evolution of $w(a)$ to converge to a common potential-independent behavior.
This is evident from \fig{F_X} where we plot the $w(a)$ behavior
for two different potentials, $V\(\phi\)=A e^{-\phi^2/\sigma^2}$
and $V\(\phi\)=A\phi^{-\alpha}$, and in both cases we choose  $F\(X\)=X_m+\(X-X_m\)^2$.
We take initial conditions $\phi_i=1$ and $X=X_m$ and adjust the constants appropriately.
The $w(a)$ behaviors turn out to be very different, indicating that there is no common
model-independent behavior for this class of models.  On the other hand, our analytic
approximation (equation \ref{Xexpand}) does give excellent agreement, although of course
the approximation is itself a function of $V(\phi)$ in this case.

\section{Observational Constraints in the $(w_0,K)$ Plane}

The results of this paper, combined with previous studies, indicate 
that Eq. (\ref{finalfinal}) applies both to quintessence models and 
to a subset of k-essence models with $w\simeq -1$. Hence Eq. (\ref{finalfinal}) 
is a useful and physically well-motivated parametrization for $w(a)$ that 
can be compared with the observations. 
So, in this section, we present the observational constraints on 
the equation of state parameters $w_0$ and on $K$.

First, we note that the cosmological constant corresponds to a {\it line} in 
the $(w_0,K)$ plane: $w_0=-1$ {\it irrespective of}  $K$. This can be understood for a canonical 
scalar field by noting that 
$w_0=-1$ corresponds to the case where the scalar field sits at the minimum $(K<1)$ or the 
maximum $(K>1)$ of the potential. 

As observational data we consider the recent compilation of 397 Type Ia supernovae (SNIa), 
  called the Constitution set with the light curve fitter SALT, by Hicken et al. \cite{hicken} and 
the measurements of baryon acoustic oscillations (BAO) from the SDSS data \cite{bao}.
\footnote{We do not consider BAO distance measurements by Percival 
et al. \cite{percival} because some points of tension were noted between 
the data sets \cite{percival}.}
Uncertainties in the distance modulus of a supernova include uncertainties in 
light curve fitting parameters (the maximum magnitude, stretch parameter, color 
correction parameter) and due to the peculiar velocity (400 ${\rm km s^{-1}}$) 
as given in \cite{hicken}.

BAO  measurements from the SDSS data provide a constraint on the distance parameter 
 $A$ defined by
\beqa
A(z)=(\Om_mH_0^2)^{1/2}\left(\frac{1}{H(z)z^2}\int^z_0\frac{dz'}{H(z')}\right)^{2/3}
\eeqa
to be $A(z=0.35)=0.469\pm0.017$.\footnote{Note that $A$ changes only slightly in varying 
the spectral index of the matter power spectrum, $n$. 
This value is for $n=0.98$, which changes to $A=0.472$ if $n=0.96$. }

\begin{figure}
\epsfig{file=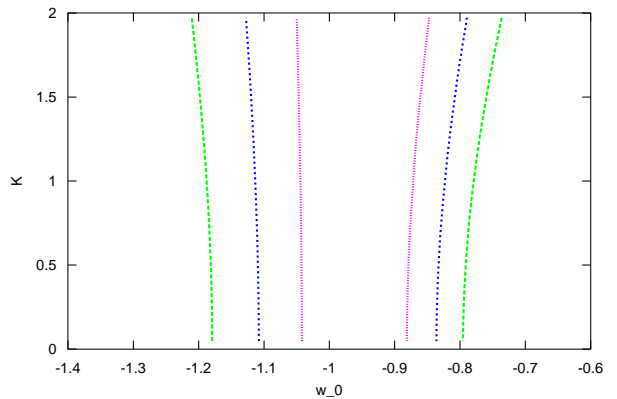,height=70mm}
\caption{Contours at 68.3\% (red, inner), 95.4\% (blue, middle), 99.7\% (green, outer) confidence 
level on $w_0$ and $K$. The Constitution SN set 
was combined with BAO constraint. }	
\label{sndata}
\end{figure}

The joint constraints from SNIa and BAO are shown in Fig. \ref{sndata}. We marginalize over $\Om_m$ 
to derive the constraints. 
The allowed range of $w_0$ is narrow: $-1.04\siml w_0\siml -0.86 (1 \sigma)$. 
\footnote{We note that SNIa data alone do not constrain $w_0$ much; $-1.2\siml w_0\siml -0.7$.} 
We find that the cosmological constant $w_0=-1$ is fully consistent with the current data. 
Note that $K$, which parametrizes the curvature of $V(\phi_i)$, is not well-constrained by 
current SNIa and BAO data. 

\section{Conclusions}

Our results indicate that k-essence models with $w$ near $-1$ can be divided into two broad categories:
models with $X \approx 0$ and those with $F_X \approx 0$.  In the former case, we find a generic
evolution for $w(a)$ which is identical to the previously-derived evolution for quintessence.  This
strengthens the case that equation (\ref{finalfinal}) is a useful and physically well-motivated
parametrization for $w(a)$ that can be compared with the observations, since it applies both to
quintessence models and to a subset of k-essence models with $w \approx -1$. 
Applying this parametrization to SNIa data and BAO, we find that the present-day value of $w$
is constrained to lie near $-1$, while the curvature parameter $K$ is poorly constrained by the observations. 
Further, we see that the cosmological constant limit of these models is consistent 
with the current data. 

On the other hand, k-essence models with $F_X \approx 0$ can demonstrate quite different behavior.
In this case, the evolution of $w(a)$ is strongly dependent on the particular potential, and there is
no ``generic" behavior.

\section*{Acknowledgments}
T.C. was supported in part by a Grant-in-Aid for Scientific Research
from JSPS (No.\,20540280)
and from MEXT (No.\,20040006) and in part by Nihon University. 
S.D. and R.J.S. were supported in part by the Department of
Energy (DE-FG05-85ER40226). Some of numerical computations were 
performed at YITP in Kyoto University.




\end{document}